\documentclass[aps,preprint,showpacs,superscriptaddress,groupedaddress,nofootinbib]{revtex4}  
\usepackage{graphicx}  
\usepackage{dcolumn}   
\usepackage{bm}        
\usepackage{amssymb}   
\usepackage{amsmath}
\usepackage{amsfonts}

\begin{document}



\title{The Gauged Thirring Model in Thermodynamic Equilibrium}
%
\author{C. A. Bonin} \affiliation{Rua Cl\'{a}udia Pietrobelli Mongruel, 107 (lado). Bairro: Brasilinha. Pira\'{i} do Sul, PR. Brazil. CEP: 84240-000 \\ carlosbonin@gmail.com (corresponding author)}
\author{B. M. Pimentel} \affiliation{ S\~{a}o Paulo State University - Institute for Theoretical Physics (IFT/UNESP), Rua Dr. Bento Teobaldo Ferraz 271 - Bl. II - Barra Funda CEP 01140-070 - S\~{a}o Paulo, S\~{a}o Paulo - Brazil\\%
pimentel@ift.unesp.br }


\date{\today}

\begin{abstract}
We study the Gauged Thirring Model (also known as Kondo Model) in thermodynamic equilibrium using the Matsubara-Fradkin-Nakanishi formalism. In this formulation, both the temperature and the chemical potential are kept to be nonvanishing. Starting from the field equations, we write down the Dyson-Schwinger-Fradkin equations, the Ward-Fradkin-Takahashi identities, and expressions for the thermodynamical generating functional. We find the partition function of the theory and  study some key features of its exact two-point Green Functions, including the Landau-Khalatnikov/Fradkin transformations and some limiting cases of interest as well. In particular, we show that we can recover results from both the Schwinger and the Thirring's models from the Kondo Model in thermodynamic equilibrium.
\end{abstract}

\maketitle


\section{Introduction}

Few examples of Quantum Field Theories are known to be exactly solvable at vanishing temperature \cite{Strocchi, Abdalla}. Among those, the Thirring, the Schwinger, and the Kondo models are textbook examples. The Thirring Model consists of massless Fermion fields interacting with on another in a manner - in spite of the number of flavors - quite similar to Fermi's interaction for beta decay, with two major differences: the global symmetry is Abelian and the spacetime is two dimensional \cite{Thirring, Fermi}. The Schwinger Model is a two dimensional, massless version of quantum electrodynamics \cite{Schwinger}. The Kondo Model is, essentially, the Thirring Model reformulated with a local Abelian symmetry and, for that reason, the model is known as the Gauged Thirring Model \cite{Kondo1}.

All three of these models are two dimensional, which means they barely  bear  a resemblance (if they do at all) with the physical world. On that subject, Thirring himself wrote in the paper in which he presented his model ``\textit{The merit this model may have is more of pedagogical nature since it shows explicitly what a relativistic theory can look like}" \cite{Thirring}. And, as a matter of fact, these three models have been extensively used throughout the decades since they first appeared  as abstract testing grounds in which better understanding of quantum fields can be attained. For instance, soon after the Thirring Model was proposed, Johnson studied some traits of the model's exact Green Functions, like  their associated infrared problem and the definition of products of singular field operators at the same spacetime points \cite{Johnson}. Sommerfield used the Thirring Model and a generalization of it which consists of the Thirring's model interacting with a Boson field in order to study the definition of currents in two dimensions \cite{Sommerfield}. A more formal treatment for operator solutions of the Schwinger model was given by Lowenstein and Swieca \cite{Lowenstein}. Looking into a massive, non-Abelian generalization of the Schwinger model, Coleman, Jackiw, and Susskind investigated quark confinement and charge shielding \cite{Coleman}. Coleman also showed the equivalence between the massive Thirring Model and the sine-Gordon model \cite{S Coleman}. On the lines of a non-Abelian generalization of the Schwinger Model, Arod\'{z} examined the quark Green function and the Fermionic determinant in quantum chromodynamics in two dimensions \cite{Arodz}. A decade later than Lowenstein and Swieca, Capri and Ferrari extended their formal work with the Schwinger Model to include analysis of the chiral anomaly \cite{Capri}. Soldate used the Schwinger Model to investigate whether assumptions made in a nonperturbative technique known as Operator Product Expansion hold true on a model that could be solved exactly \cite{Soldate}. Topology was also focus of research in exactly solvable models, with the study of the Schwinger Model on a sphere by Jayewardena being one of the pioneer works \cite{Jayewardena}. Also noteworthy is Bardakci and Crescimanno's study of the Schwinger Model on the plane with emphasis on the topological aspects of the gauge field \cite{Bardakci}. Sachs and Wipf studied the chiral symmetry breaking in some two dimensional gauge theories, including the Schwinger Model \cite{Sachs Wipf Temp and Curv}. $n$-point correlation functions for the Schwinger model were calculated by Steele, Subramanian, and Zahed \cite{Steele}. Kondo himself studied the mapping between the Fermions in the massive Gauged Thirring model and Bosons, a process known as Bosonization \cite{Kondo2}. Bosonization was also studied in the massive Thirring model in arbitrary dimensions by Ikegami, Kondo, and Nakamura \cite{Kondo3}. The Gauged Thirring Model was further studied in (2+1) dimensions in the Heisenberg picture and with the Causal Perturbation Theory by one of us and collaborators \cite{Manzoni1999, Lunardi, Manzoni2000}. Going back to two dimensions, using path-integral techniques, Bosonization was researched in the massive Gauged Thirring Model by one of us and a collaborator \cite{Bufalo Path Integral}. Furthermore, one of us and collaborators showed that, at quantum level, both the Thirring Model and the Schwinger Model are limiting cases of the Gauged Thirring Model \cite{Bufalo Strong Limits}. This last result lies at the core of as to why we choose, in the present paper, to work with the Gauged Thirring Model: from the Gauged Thirring Model we can draw conclusions about not only the Kondo Model itself, but also both the Thirring and the Schwinger's models, at least in the zero temperature scenario.

Besides giving rise to fascinating phenomena like symmetry breaking restoration, thermal masses, and Debye screening, just to cite a few, finite temperature effects  pose additional challenges to computation of physical quantities in  Quantum Field Theories, \textit{exempli gratia}, the technical problem of evaluating Matsubara sums and the breakdown of the na\"{i}ve perturbation theory \cite{Le Bellac, Das, Kapusta}. As such, it is not surprising that solvable models have been subject of research of typical Quantum Field Theory at Finite Temperature's processes and techniques. One of the first works to take effects of temperature into account in solvable field models was the computation of the photon and the electron propagators in the finite-temperatured Schwinger Model by Stam and Visser \cite{Stam}. Ruiz Ruiz and Alvarez-Estrada sought exact solutions for the Schwinger and the Thirring models at finite temperature and studied the thermodynamics of those models as well \cite{Ruiz Ruiz Alvarez Estrada Exact Schwinger, Ruiz Ruiz Alvarez Estrada Exac Thirring, Ruiz Ruiz Alvarez Estrada Thermo}. Later, Alvarez-Estrada and Nicola included finite chemical potential in their analysis of those models \cite{Alvarez Estrada Nicola}. In their work about chiral symmetry breaking in two dimensional field theories, Sachs and Wipf studied how both the curvature and the temperature affect the phenomenon \cite{Sachs Wipf Temp and Curv}. Sachs and Wipf also studied thermodynamics and conformal properties of generalizations of the Thirring Model and, using the Euclidean Path-Integral Technique, calculated the temperature dependence of the order parameter in the Schwinger Model \cite{Sachs, Sachs Finite T Schwinger}.

In the present paper we study the Gauged Thirring Model in thermodynamic equilibrium. As mentioned before, we deal in particular with the Kondo Model because, at vanishing temperature, this model was shown to be a generalization of both the Thirring and the Schwinger's models. Our primary goals are to investigate whether such a generalization holds true at finite temperature and study the thermodynamic properties of the model as well. In order to achieve our goals, we use the Matsubara-Fradkin-Nakanishi formalism. The origins of this formalism date back to the original work of Matsubara in the canonical ensemble for nonrelativistic systems \cite{Matsubara}. Fradkin extended Matsubara's method to deal with relativistic theories in the grand canonical ensemble \cite{Fradkin}. At last, in order to deal with gauge theories in a Lorentz covariant way, we invoke Nakanishi's Auxiliary Field Method \cite{Nakanishi 1, Nakanishi and Ojima}. To the collection of these three approaches we call ``Matsubara-Fradkin-Nakanishi Formalism". A constructive presentation of this formalism can be found in a previous work of ours, followed by its application to a gauge theory in thermodynamic equilibrium \cite{Bonin}.

This paper is organized as follows. In section \ref{The Gauged Thirring Model in the Matsubara-Fradkin Formalism} we present the Gauged Thirring Model and we use the Matsubara-Fradkin-Nakanishi Formalism in order to achieve a quantum version of the theory in thermodynamic equilibrium. In section \ref{The Thermodynamical Generating Functional and General Properties of the Green Functions} we study the thermodynamical generating functional for the theory and analyse some general properties of its complete Green Functions, including the Dyson-Schwinger-Fradkin equations, the Ward-Fradkin-Takahashi identities, and the Landau-Khalatnikov/Fradkin transformations. In section \ref{The Green Functions and the Partition Function} we write down the complete two-point Green functions of the model, inspect several limits of them, and compute the partition function. Our final remarks are presented in section \ref{Final Remarks}. An appendix is added concerning the most general form of the inverse of a rank-two tensor which has the properties of the gauge field Green Function in thermodynamic equilibrium in two dimensions. Throughout this paper we use Einstein's implicit summation convention and the natural system of units in which the Planck constant, the speed of light, and the Boltzmann constant all have unit values.

\section{The Gauged Thirring Model in the Matsubara-Fradkin-Nakanishi Formalism}\label{The Gauged Thirring Model in the Matsubara-Fradkin Formalism}

In this section we present the fundamentals of the study of the Gauged Thirring Model in thermodynamic equilibrium. The Gauged Thirring Model was introduced by Kondo \cite{Kondo1} and it consists in the following Lagrangian density in $(1+1)$ dimensions:

\begin{equation}
\mathcal{L}=-\overline{\psi}_{a}i\left(\gamma^{\mu}\right)_{ab}\partial_{\mu}\psi_{b} +\left(\gamma^{\mu}\right)_{ab}\mathcal{A}_{\mu}\overline{\psi}_{a}\psi_{b} +\frac{1}{2g}\left(\mathcal{A}_{\mu}-\partial_{\mu}\theta\right)\left(\mathcal{A}^{\mu}-\partial^{\mu}\theta\right) -\frac{1}{4q^{2}}\mathcal{F}_{\mu\nu}\mathcal{F}^{\mu\nu}.\label{Lagrangian}
\end{equation}

Here, $\overline{\psi}$ and $\psi$ are Grassmannian fields, $\gamma$'s are matrices satisfying\footnote{Throughout this paper we use the definitions $\left\{ A,B\right\}\equiv AB+BA$ for the anticommutator and $\left[A,B\right]\equiv AB-BA$ for the commutator. Attention should be paid in order to avoid confusion between the anticommutator and the set with two elements: their notations are the same.}

\begin{equation}
  \left\{\gamma^\mu ,\gamma^\nu\right\}_{ab} = 2\delta_{ab}\eta_M^{\mu\nu}
\end{equation}
where $a,b\in\left\{1,2\right\}$ are Dirac indices, $\mu,\nu\in\left\{0,1\right\}$ are spacetime indices, and $\eta_M$ is the Minkowski metric tensor with contravariant diagonal $\mbox{diag}\left(\eta_M\right)=\left[1,-1\right]$, $\mathcal{A}$ is the $U\left(1\right)$ gauge field (henceforth sometimes called the electromagnetic field), $g$ is a dimensionless parameter (associated with the Thirring Model's coupling constant), $\theta$ is the St\"{u}ckelberg's auxiliary field \cite{Stueckelberg, Stueckelberg Field}, $q$ is a constant with dimension of energy (related to the Schwinger Model's coupling constant), and finally $\mathcal{F}_{\mu\nu}\equiv\partial_\mu\mathcal{A}_\nu-\partial_\nu\mathcal{A}_\mu$ are the components of the field-strength.

Due to the invariance under local $U(1)$ gauge transformation,

\begin{eqnarray}
\psi_{a}\left(x\right) & \rightarrow & \psi_{a}^{\prime}\left(x\right)=e^{i\lambda\left(x\right)}\psi_{a}\left(x\right);\label{U1 1}\\
\overline{\psi}_{a}\left(x\right) & \rightarrow & \overline{\psi}_{a}^{\prime}\left(x\right)=\overline{\psi}_{a}\left(x\right)e^{-i\lambda\left(x\right)};\\
\mathcal{A}_{\mu}\left(x\right) & \rightarrow & \mathcal{A}_{\mu}^{\prime}\left(x\right)=\mathcal{A}_{\mu}\left(x\right)+\partial_{\mu}\lambda\left(x\right);\\
\theta\left(x\right) & \rightarrow & \theta^{\prime}\left(x\right)=\theta\left(x\right)+\lambda\left(x\right),\label{U1 2}
\end{eqnarray}
for an arbitrary Lorentz scalar field  $\lambda$, the model (\ref{Lagrangian}) possesses the following conserved charge:

\begin{equation}
N=\int dx\left(\gamma^{0}\right)_{ab}\overline{\psi}_{a}\left(x\right)\psi_{b}\left(x\right).\label{conserved charge}
\end{equation}

Following Nakanishi's auxiliary-field method, we postulate a Lagrangian
density operator $\widehat{\mathcal{L}}_{N}$ to describe the quantum
theory in thermodynamic equilibrium:
\begin{eqnarray}
\widehat{\mathcal{L}}_{N} & \equiv & \frac{1}{2}\left(\gamma_{\mu}^{E}\right)_{ab}\left[\widehat{\overline{\psi}}_{a}^{s}, \partial_{\mu}^{\left(\mu_{f}\right)}\widehat{\psi}_{b}^{s}\right] +\frac{i}{2}\left(\gamma_{\mu}^{E}\right)_{ab}\widehat{A}_{\mu}^{s}\left[\widehat{\overline{\psi}}_{a}^{s},\widehat{\psi}_{b}^{s}\right]+\nonumber \\
 &  & -\frac{1}{2g}\left(\widehat{A}_{\mu}^{s}-\partial_{\mu}\widehat{\theta}^{s}\right)\left(\widehat{A}_{\mu}^{s}-\partial_{\mu}\widehat{\theta}^{s}\right) -\frac{1}{4q^{2}}\widehat{F}_{\mu\nu}^{s}\widehat{F}_{\mu\nu}^{s}+\frac{1}{2}\left\{ \widehat{B},\widehat{G}\left[\widehat{A}^{s}\right]\right\} +\frac{\alpha}{2}\widehat{B}^{2}+\nonumber \\
 &  & +\frac{1}{2}\left[\overline{\eta}_{a},\widehat{\psi}_{a}^{s}\right]+\frac{1}{2}\left[\eta_{a},\widehat{\overline{\psi}}_{a}^{s}\right] +J_{\mu}\widehat{A}_{\mu}^{s}+\varsigma\widehat{\theta}^{s}.\label{enjoo}
\end{eqnarray}
where $\bar{\eta}$ and $\eta$ are classical Grassmannian field sources,  $J$ is a classical $SO\left(2\right)$  vector  field source,  $\varsigma$ is a classical real $SO\left(2\right)$ scalar field source and, for any field $\widehat{\phi}$:\footnote{For the dependence of a field $\widehat{\phi}$ we can use any of the following notations, as convenience dictates, $\widehat{\phi}\left(x\right)=\widehat{\phi}\left(x_1,x_0\right)=\widehat{\phi}\left(x_1,\tau_x\right)$, where the variable $x_0=\tau_x$ is the one associated with the temperature.}
\begin{equation}
\widehat{\phi}^{s}\left(x_1,\tau_x\right)\equiv \widehat{\rho}_{s}^{-1}\left(\tau_x\right)\widehat{\phi}\left(x_1,0\right)\widehat{\rho}_{s}\left(\tau_x\right),\label{with the first density matrix}
\end{equation}
where $\widehat{\rho}_s$ is the density matrix
\begin{equation}
\widehat{\rho}_{s}\left(\beta\right)=e^{-\beta\left(\widehat{H}_{T}^{\prime}+\mu_{f}\widehat{N}\right)},\label{first density matrix}
\end{equation}
with $\beta$ being the inverse of the temperature $T$, $\widehat{H}_{T}^{\prime}$ being the Hamiltonian (which includes
the classical sources), and $\mu_{f}$ being a Lagrangian multiplier (henceforth called (Fermionic) chemical potential) associated with the quantum version of the conserved charge (\ref{conserved charge}),
\begin{equation}
\widehat{N}=\frac{\left(\gamma^{0}\right)_{ab}}{2}\int_{V}dx\left[\widehat{\overline{\psi}}_{a}\left(x\right),\widehat{\psi}_{b}\left(x\right)\right],
\end{equation}
multiplied by the negative of the inverse of the temperature.
We also use the notation
\begin{equation}
\partial_{\mu}^{\left(\mu_{f}\right)}\equiv\partial_{\mu}+\mu_{f}\delta_{\mu0}.
\end{equation}

$\widehat{B}$ in (\ref{enjoo}) is called the Nakanishi's auxiliary field, $\alpha$ is a non vanishing real parameter (the gauge parameter), $\widehat{G}$ is the
gauge choice operator - which has the following property: under a gauge
transformation, $\widehat{G}$ is \textit{not}, in general,
invariant: $\widehat{G}\left[\widehat{A}\right]\rightarrow\widehat{G}\left[\widehat{A}^{\prime}\right]\neq\widehat{G}\left[\widehat{A}\right]$. For simplicity, we choose to work with the $R_\alpha$ condition $\widehat{G}\left[\widehat{A}\right]=\widehat{G}_{R_{\alpha}}\left[\widehat{A},\widehat{\theta}\right]$:\footnote{The $R_\alpha$ condition ensures that the field equations involving the gauge field (\ref{For the gauge field})  and  the St\"{u}ckelberg's auxiliary field (\ref{Stuckelberg}) do not mix with each other.}

\begin{equation}
\widehat{G}_{R_{\alpha}}\left[\widehat{A}^{s},\widehat{\theta}^{s}\right]\equiv\frac{1}{q}\partial_{\mu}\widehat{A}_{\mu}^{s}-\alpha\frac{q}{g}\widehat{\theta}^{s}.\label{alfa}
\end{equation}

From equation (\ref{enjoo}) onward all the implicit sums $a\cdot b\equiv a_\mu b_\mu$ for any vectors $a$ and $b$ are performed using the Euclidean metric and
we have the Euclidean Dirac matrices $\gamma^E$'s as well, satisfying:
\begin{equation}
\left\{\gamma_{\mu}^{E},\gamma_{\nu}^{E}\right\}_{ab}=2\delta_{ab}\delta_{\mu\nu}.\label{Euclidean Dirac}
\end{equation}

Using the Schwinger's principle, we find our first set of field equations \cite{Schwinger Principle}:

\begin{eqnarray}
\left(\gamma_{\mu}^{E}\right)_{ab}\widehat{D}_{\mu}^{\left(\mu_{f}\right)}\left[\widehat{A}^s\right]\widehat{\psi}_{b}^{s}\left(x\right) & = & \eta_{a}\left(x\right)\widehat{1};\label{first of the set}\\
\left(\gamma_{\mu}^{E}\right)_{ba}\widehat{D}_{\mu}^{\left(-\mu_{f}\right)}\left[-\widehat{A}^s\right]\widehat{\overline{\psi}}_{b}^{s}\left(x\right) & = & \overline{\eta}_{a}\left(x\right)\widehat{1};\\
P_{\mu\nu}\widehat{A}_{\nu}^{s}\left(x\right) & = & \frac{i}{2}\left(\gamma_{\mu}^{E}\right)_{ab}\left[\widehat{\overline{\psi}}_{a}^{s}\left(x\right),\widehat{\psi}_{b}^{s}\left(x\right)\right] +J_{\mu}\left(x\right)\widehat{1};\label{For the gauge field}\\
\frac{1}{g}\left(\Delta+\alpha\frac{q^{2}}{g}\right)\widehat{\theta}^{s}\left(x\right) & = & \varsigma\left(x\right)\widehat{1}\label{Stuckelberg};
\end{eqnarray}
where $\widehat{D}_{\mu}^{\left(\mu_{f}\right)}\left[\widehat{A}\right] \equiv  \widehat{1}\partial_\mu^{\left(\mu_f\right)}+i\widehat{A}_\mu$, $\Delta \equiv -\partial_\mu\partial_\mu$ is the negative of the Laplace operator, and

\begin{equation}
P_{\mu\nu}  \equiv  \left(\frac{1}{q^{2}}\Delta+\frac{1}{g}\right)\delta_{\mu\nu}+\frac{1}{q^{2}}\left(1-\frac{1}{\alpha}\right)\partial_{\mu}\partial_{\nu}. \label{P}
\end{equation}

In writing this set of equations, we have already solved exactly the equation for the Nakanishi's auxiliary field and inserted the result in (\ref{For the gauge field}).

In the absence of the external classical sources, the Lagrangian density
operator (\ref{enjoo})  is invariant under a quantum version of the $U\left(1\right)$ gauge transformation (\ref{U1 1}-\ref{U1 2}) - that is, with operators instead of classical fields and with the additional rule $\widehat{B}\left(x\right)  \rightarrow  \widehat{B}^{\prime}\left(x\right)=\widehat{B}\left(x\right)$ -
provided the auxiliary field operator  vanishes identically. 
If $\widehat{B}\left(x\right)\neq\widehat{0}$, on the other hand,
the gauge choice operator breaks the gauge symmetry of the theory
for a general parameter operator $\widehat{\lambda}$.
However, if the parameter operator is carefully chosen in such a way that it makes the gauge
choice operator invariant, then the theory is again gauge invariant
for that specific choice of the parameter operator. So, for the
$R_{\alpha}$ gauge choice (\ref{alfa}), the invariance of the gauge choice operator under a gauge transformation, \textit{i.e.}, $\widehat{G}_{R_{\alpha}}\left[\widehat{A}^{s}+\partial\widehat{\Lambda},\widehat{\theta}^{s}+\widehat{\Lambda}\right]= \widehat{G}_{R_{\alpha}}\left[\widehat{A}^{s},\widehat{\theta}^{s}\right]$, implies:

\begin{equation}
\left(\Delta+\alpha\frac{q^{2}}{g}\right)\widehat{\Lambda}\left(x\right)=\widehat{0}.\label{pastel}
\end{equation}

In order to preserve the gauge invariance of the model at quantum level, we add the relation above  as a constraint to the Lagrangian density operator (\ref{enjoo}). We do so by writing $\widehat{\varkappa}\left(x\right)\equiv i\kappa\widehat{\overline{C}}\left(x\right)\upsilon$ as a Lagrange multiplier operator with $\kappa$ being a constant to be fixed \textit{a posteriori} and $\upsilon$ being a Grassmannian constant. We also define the field $\widehat{C}\left(x\right)\equiv\upsilon\widehat{\Lambda}\left(x\right)$. $\widehat{C}$ and $\widehat{\overline{C}}$ are called ghost field operators. After including classical (Grassmannian) sources for the ghost fields, we end up with a new Lagrangian density operator:

\begin{equation}
\widehat{\mathcal{L}}_{gs}=\widehat{\mathcal{L}}_{N}+i\kappa\widehat{\overline{C}}\left(\Delta+\alpha\frac{q^{2}}{g}\right)\widehat{C} +\frac{1}{2}\left[\overline{d},\widehat{C}^{s}\right]+\frac{1}{2}\left[d,\widehat{\overline{C}}^{s}\right].
\end{equation}


The field equations for the ghost fields are found in the same way
the others were:
\begin{eqnarray}
i\kappa\left(\Delta+\alpha\frac{q^{2}}{g}\right)\widehat{C}^{s}\left(x\right) & = & d\left(x\right)\widehat{1};\label{ghost C}\\
i\kappa\left(\Delta+\alpha\frac{q^{2}}{g}\right)\widehat{\overline{C}}^{s}\left(x\right) & = & -\overline{d}\left(x\right)\widehat{1},\label{ghost C bar}
\end{eqnarray}
and this completes our set of field equations for the model.

Due to the presence of the ghost fields, in the absence of the external
sources, the theory has a new global internal symmetry:\footnote{Fields written without the subscript $[s]$ correspond to fields without the classical sources.}
\begin{eqnarray}
\widehat{C}\left(x\right) & \rightarrow & \widehat{C}^{\prime}\left(x\right)=e^{i\theta_{0}}\widehat{C}\left(x\right);\\
\widehat{\overline{C}}\left(x\right) & \rightarrow & \widehat{\overline{C}}^{\prime}\left(x\right)=\widehat{\overline{C}}\left(x\right)e^{-i\theta_{0}},
\end{eqnarray}
where $\theta_0\in\mathbb{R}$. Thanks to this symmetry, there is another Noether
charge in the problem:
\begin{equation}
\widehat{Q}=\frac{1}{2}\int_V dx_1\left\{ \left[\widehat{\overline{\pi}}\left(x_1,\tau_x\right),\widehat{C}\left(x_1,\tau_x\right)\right] +\left[\widehat{\overline{C}}\left(x_1,\tau_x\right),\widehat{\pi}\left(x_1,\tau_x\right)\right]\right\},
\end{equation}
where $V$ is the ``one-dimensional volume", that is, the total length of the thermodynamical system which, without loss of generality for our purposes, is infinite, $\widehat{\overline{\pi}}$ and $\widehat{\pi}$
are the conjugated canonical momentum operators to $\widehat{C}$
and $\widehat{\overline{C}}$, respectively.
$\widehat{Q}$ is called the ghost charge. Since we have found a new conserved quantity in process of quantization, the density matrix of the problem is redefined to be
\begin{equation}
\widehat{\rho}_{gs}\left(\beta\right)=e^{-\beta\left(\widehat{H}_{T}-\mu_{f}\widehat{N}-\mu_{g}\widehat{Q}\right)},
\end{equation}
where  $\widehat{H}_{T}$ is the total Hamiltonian (which includes
the ghost fields and all the sources) and $\beta\mu_{g}$ is the Lagrange multiplier associated with the ghost charge ($\mu_g$ is called the ghost chemical
potential).

For any field $\widehat{\phi}$, we define:
\begin{equation}
\widehat{\phi}^{gs}\left(x_{1},\tau_{x}\right)\equiv\widehat{\rho}_{gs}^{-1}\left(\tau_x\right)\widehat{\phi}\left(x_{1},0\right) \widehat{\rho}_{gs}\left(x_{1},\tau_{x}\right).
\end{equation}

Let $\widehat{w}$ be a field that commutes with the ghost fields.
We can show that
\begin{equation}
\widehat{w}^{gs}\left(x_{1},\tau_{x}\right)=\widehat{w}^{s}\left(x_{1},\tau_{x}\right).
\end{equation}

On the other hand, for the ghost fields themselves,
\begin{eqnarray}
\widehat{C}^{gs}\left(x_{1},\tau_{x}\right) & = & e^{\tau_x\mu_{g}}\widehat{C}^{s}\left(x_{1},\tau_{x}\right);\\
\widehat{\overline{C}}^{gs}\left(x_{1},\tau_{x}\right) & = & -e^{\tau_x\mu_{g}}\widehat{\overline{C}}^{s}\left(x_{1},\tau_{x}\right).
\end{eqnarray}

In these cases, we understand that fields with subscript $s$ are obtained through equations (\ref{with the first density matrix}) and (\ref{first density matrix}).

As we know, for any two fields:
\begin{eqnarray}
\frac{\delta\widehat{\rho}_{gs}\left(\beta\right)}{\delta s_{a}\left(x_{1},\tau_{x}\right)} & = & \widehat{\rho}_{gs}\left(\beta\right)\widehat{\phi}_{a}^{s}\left(x_{1},\tau_{x}\right);\\
\frac{\delta^{2}\widehat{\rho}_{gs}\left(\beta\right)}{\delta s_{b}\left(y_{1},\tau_{y}\right)\delta s_{a}\left(x_{1},\tau_{x}\right)} & = & \widehat{\rho}_{gs}\left(\beta\right)T\left[\widehat{\phi}_{b}^{s}\left(y_{1},\tau_{y}\right)\widehat{\phi}_{a}^{s}\left(x_{1},\tau_{x}\right)\right];
\end{eqnarray}
where $s_{a}\left(x_{1},\tau_{x}\right)$ is the source for the field
$\widehat{\phi}_{a}^{s}\left(x_{1},\tau_{x}\right)$ and\footnote{The plus sign in the definition of the operation $T$ refers to non-Grassmmannian fields, whereas the minus sign is used for Grassmmannian variables.}
\begin{equation}
T\left[\widehat{A}\left(\tau_{x}\right)\widehat{B}\left(\tau_{y}\right)\right]\equiv\left\{ \begin{array}{c}
\widehat{A}\left(\tau_{x}\right)\widehat{B}\left(\tau_{y}\right),\text{ \ if }\tau_{x}=\tau_{y};\\
\theta\left(\tau_{x}-\tau_{y}\right)\widehat{A}\left(\tau_{x}\right)\widehat{B}\left(\tau_{y}\right)\pm\theta\left(\tau_{y}-\tau_{x}\right)\widehat{B}\left(\tau_{y}\right)\widehat{A}\left(\tau_{x}\right)\text{, otherwise.}
\end{array}\right.
\end{equation}

Besides, we have the grand canonical partition function
\begin{equation}
Z\left(\beta,\mu_f\right)=\text{Tr}\left[\widehat{\rho}_{g}\left(\beta\right)\right],
\end{equation}
where $\widehat{\rho}_{g}\left(\beta\right)=\left.\widehat{\rho}_{gs}\left(\beta\right)\right\vert _{s=0}$ is the density matrix without external sources and, for any operator $\widehat{F}$, we define its ensemble average $\left\langle \widehat{F}\right\rangle$ as:
\begin{equation}
\left\langle \widehat{F}\right\rangle \equiv\frac{\text{Tr}\left[\widehat{\rho}_{g}\left(\beta\right)\widehat{F}\right]}{Z\left(\beta,\mu_f\right)}.\label{thermo expect field}
\end{equation}

We define still the thermodynamical generating functional as:
\begin{equation}
Z_{GF}\left[s\right]\equiv\text{Tr}\left[\widehat{\rho}_{gs}\left(\beta\right)\right].
\end{equation}

Clearly:
\begin{equation}
Z_{GF}\left[0\right]=Z\left(\beta,\mu_f\right)\label{part func}
\end{equation}
and
\begin{eqnarray}
\left\langle \widehat{\phi}_{a}\left(x_{1},\tau_{x}\right)\right\rangle  & = & 
\frac{1}{Z\left(\beta,\mu_f\right)}\left.\frac{\delta Z_{GF}\left[s\right]}{\delta s_{a}\left(x_{1},\tau_{x}\right)}\right\vert _{s=0};\\
\left\langle T\left[\widehat{\phi}_{b}\left(y_{1},\tau_{y}\right)\widehat{\phi}_{a}\left(x_{1},\tau_{x}\right)\right]\right\rangle  & = & 
\frac{1}{Z\left(\beta,\mu_f\right)}\left.\frac{\delta^{2}Z_{GF}\left[s\right]}{\delta s_{b}\left(y_{1},\tau_{y}\right)\delta s_{a}\left(x_{1},\tau_{x}\right)}\right\vert _{s=0}. \label{two points}
\end{eqnarray}

It is also possible to show that quantities like (\ref{two points}), $\left\langle T\left[\widehat{\phi}_{b}\left(y\right)\widehat{\phi}_{a}\left(x\right)\right]\right\rangle$, depend only on the difference of the points $x-y$ \cite{Strocchi}.

\section{The Thermodynamical Generating Functional and General Properties of the Green Functions}\label{The Thermodynamical Generating Functional and General Properties of the Green Functions}

In this section we study general properties of the two-point Green Functions and of the thermodynamical generating functional for the Kondo Model.

By multiplying each field equation (\ref{first of the set}-\ref{Stuckelberg},\ref{ghost C},\ref{ghost C bar}) by the density matrix and taking the trace, we find the set of functional equations satisfied by the thermodynamical generating functional:

\begin{eqnarray}
\left(\gamma_{\mu}^{E}\right)_{ab}\partial_{\mu}^{\left(\mu_{f}\right)}\frac{\delta Z_{GF}\left[s\right]}{\delta\overline{\eta}_{b}\left(x\right)} & = & -i\left(\gamma_{\mu}^{E}\right)_{ab}\frac{\delta^{2}Z_{GF}\left[s\right]}{\delta J_{\mu}\left(x\right)\delta\overline{\eta}_{b}\left(x\right)}+\eta_{a}\left(x\right)Z_{GF}\left[s\right];\label{psi}\\
\left(\gamma_{\mu}^{E}\right)_{ba}\partial_{\mu}^{\left(-\mu_{f}\right)}\frac{\delta Z_{GF}\left[s\right]}{\delta\eta_{b}\left(x\right)} & = & i\left(\gamma_{\mu}^{E}\right)_{ba}\frac{\delta^{2}Z_{GF}\left[s\right]}{\delta J_{\mu}\left(x\right)\delta\eta_{b}\left(x\right)}+\overline{\eta}_{a}\left(x\right)Z_{GF}\left[s\right]; \\
P_{\mu\nu}\frac{\delta Z_{GF}\left[s\right]}{\delta J_{\nu}\left(x\right)} & = & i\left(\gamma_{\mu}^{E}\right)_{ab}\frac{\delta^{2}Z_{GF}\left[s\right]}{\delta\eta_{a}\left(x\right)\delta\overline{\eta}_{b}\left(x\right)} +J_{\mu}\left(x\right)Z_{GF}\left[s\right];\\
\frac{1}{g}\left(\Delta+\alpha\frac{q^{2}}{g}\right)\frac{\delta Z_{GF}\left[s\right]}{\delta\varsigma\left(x\right)} & = & \varsigma \left(x\right) Z_{GF}\left[s\right];\label{theta}\\
i\kappa\left(\Delta+\alpha\frac{q^{2}}{g}\right)\frac{\delta Z_{GF}\left[s\right]}{\delta\overline{d}\left(x\right)} & = & d\left(x\right)Z_{GF}\left[s\right];\label{c}\\
i\kappa\left(\Delta+\alpha\frac{q^{2}}{g}\right)\frac{\delta Z_{GF}\left[s\right]}{\delta d\left(x\right)} & = & -\overline{d}\left(x\right)Z_{GF}\left[s\right].\label{c barra}
\end{eqnarray}

Without solving this set of functional equations, we can study the Green Functions of the model. Solving them would ultimately give us the partition function (\ref{part func}), from which all thermodynamic properties derive. So, in a sense, this set of equations is the most important of the paper: all our results are encoded in it.

In order to study some properties of the Green Functions of the model, we define

\begin{eqnarray}
\left\langle T\left[\widehat{\overline{\psi}}_{b}\left(y\right)\widehat{\psi}_{a}\left(x\right)\right]\right\rangle  & \equiv & S_{ab}\left(x-y\right);\\
\left\langle T\left[\widehat{A}_{\nu}\left(y\right)\widehat{A}_{\mu}\left(x\right)\right]\right\rangle  & \equiv & D_{\mu\nu}\left(x-y\right);\\
\left\langle T\left[\widehat{\theta}\left(y\right)\widehat{\theta}\left(x\right)\right]\right\rangle  & \equiv & F\left(x-y\right);\label{F}\\
\left\langle T\left[\widehat{\overline{C}}\left(y\right)\widehat{C}\left(x\right)\right]\right\rangle  & \equiv & G\left(x-y\right).\label{G}
\end{eqnarray}

Using the definition of the left-hand side of these quantities (\ref{thermo expect field},\ref{two points}), we can show they satisfy well-defined symmetry and periodicity conditions \cite{Bonin}:

\begin{eqnarray}
S_{ab}\left(x_{1},\tau_x\right) & = & -S_{ba}\left(-x_{1},-\tau_x\right) =-S_{ab}\left(x_{1},\tau_x-\beta\right);\label{S original}\\
D_{\mu\nu}\left(x_{1},\tau_x\right) & = & D_{\nu\mu}\left(-x_{1},-\tau_x\right) = D_{\mu\nu}\left(x_{1},\tau_x-\beta\right);\label{D original}\\
F\left(x_{1},\tau_x\right) & = & i\kappa g G\left(x_{1},\tau_x\right)= F\left(-x_{1},-\tau_x\right)= F\left(x_{1},\tau_x-\beta\right).\label{F and G}
\end{eqnarray}

Incidentally, in writing these periodicity conditions, we have shown that the ghost chemical potential is a purely imaginary number: $\mu_g = i\left(2n-1\right)\pi/\beta$, where $n$ is an integer. It is also not difficult to see that $G$ is the Green function for the differential operator $i\kappa\left(\Delta+\alpha\frac{q^{2}}{g}\right)$ while $F$ is the Green function for $\frac{1}{g}\left(\Delta+\alpha\frac{q^{2}}{g}\right)$. This is the reason for the proportionality between (\ref{F}) and (\ref{G}) as stated in equation (\ref{F and G}). In order to study the other Green Functions, we define

\begin{eqnarray}
\vartheta_{\mu}\left(x\right) & \equiv & \frac{\delta \ln Z_{GF}\left[s\right]}{\delta J_{\mu}\left(x\right)};\\
\mathcal{D}_{\mu\nu}^{\left[s\right]}\left(x,y\right) & \equiv & \frac{\delta \vartheta_{\mu}\left(x\right)}{\delta J_{\nu}\left(y\right)};\label{Green photon}\\
\mathcal{S}_{ab}^{\left[s\right]}\left(x,y\right) & \equiv & \frac{\delta^2 \ln Z_{GF}\left[s\right]}{\delta\eta_{b}\left(y\right)\delta\overline{\eta}_{a}\left(x\right)}.\label{Green fermion}
\end{eqnarray}

By properly functionally deriving the equations of the set (\ref{psi}-\ref{c barra}) with respect to the classical sources, we can show that

\begin{eqnarray}
\left[\mathcal{D}_{\mu\nu}^{\left[s^{\prime}\right]}\left(x,y\right)\right]^{-1} & = & \delta\left(x_{1}-y_{1}\right)\Delta^{\left(+\right)}_\beta\left(\tau_{x}-\tau_{y}\right)P_{\mu\nu}\left(z\right) +\Pi_{\mu\nu}^{\left[s^{\prime}\right]}\left(x,y\right);\label{D inverso}\\
\left[\mathcal{S}_{ab}^{\left[s^{\prime}\right]}\left(x,y\right)\right]^{-1} & = & \delta\left(x_{1}-y_{1}\right)\Delta^{\left(-\right)}_\beta\left(\tau_{x}-\tau_{y}\right)\left(\gamma_{\mu}^{E}\right)_{ab} D_{\mu}^{\left(\mu_{f}\right)}\left[\vartheta\right]_{y} -\Sigma_{ab}^{\left[s^{\prime}\right]}\left(x,y\right),\label{S inverso}
\end{eqnarray}
where the subscript $s'$ means the vanishing of the Fermionic sources and $\Delta^{\left(+\right)}_\beta$ ($\Delta^{\left(-\right)}_\beta$) is the periodic (anti-periodic) Dirac comb distribution,

\begin{equation}
  \Delta^{\left(\pm\right)}_\beta\left(\tau\right) \equiv \sum_{n=-\infty}^{\infty} \left(\pm 1\right)^n \delta\left(\tau -n\beta\right).
\end{equation}

The objects $\Pi_{\mu\nu}^{\left[s^{\prime}\right]}$
and  $\Sigma_{ab}^{\left[s^{\prime}\right]}$ in (\ref{D inverso},\ref{S inverso}) are the components of the polarization and of the mass operators, which are given implicitly by\footnote{We use the notation $\int_{\beta V}d^2x f\left(x\right)=\int_0^\beta dx_0\int_V dx_1 f\left(x_1,x_0\right)$.}
\begin{eqnarray}
-i\left(\gamma_{\mu}^{E}\right)_{ab}\frac{\delta\mathcal{S}_{ba}^{\left[s^{\prime}\right]}\left(x,x\right)}{\delta J_{\nu}\left(y\right)} & \equiv & \int_{\beta V}d^{2}z\Pi_{\mu\xi}^{\left[s^{\prime}\right]}\left(x,z\right) \mathcal{D}_{\xi\nu}^{\left[s^{\prime}\right]}\left(z,y\right); \label{Pi}\\
-i\left(\gamma_{\mu}^{E}\right)_{ac}\frac{\delta\mathcal{S}_{cb}^{\left[s^{\prime}\right]}\left(x,y\right)}{\delta J_{\mu}\left(x\right)} & \equiv & \int_{\beta V}d^{2}z\Sigma_{ac}^{\left[s^{\prime}\right]}\left(x,z\right) \mathcal{S}_{cb}^{\left[s^{\prime}\right]}\left(z,y\right),\label{Sigma}
\end{eqnarray}
or
\begin{align}
\Pi_{\mu\nu}^{\left[s^{\prime}\right]}\left(x,y\right)= &\,\left(\gamma_{\mu}^{E}\right)_{ab}\int_{\beta V}d^{2}ud^{2}v\mathcal{S}_{bc}^{\left[s^{\prime}\right]}\left(x,v\right) \Gamma_{\nu\left(cd\right)}^{\left[s^{\prime}\right]}\left(v,u,y\right)\mathcal{S}_{da}^{\left[s^{\prime}\right]}\left(u,x\right);\\
\Sigma_{ab}^{\left[s^{\prime}\right]}\left(x,y\right)=&\, -\left(\gamma_\mu^E\right)_{ac}\int_{\beta V}d^2ud^2v \mathcal{D}_{\mu\nu}^{\left[s^{\prime}\right]}\left(x,u\right)\mathcal{S}^{\left[s^{\prime}\right]}_{cd}\left(x,v\right), \Gamma_{\nu\left(db\right)}^{\left[s^{\prime}\right]}\left(v,y,u\right)
\end{align}
where
\begin{equation}
\Gamma_{\mu\left(ab\right)}^{\left[s^{\prime}\right]}\left(x,y,z\right)\equiv-i\frac{\delta\left\{ \left[\mathcal{S}_{ab}^{\left[s^{\prime}\right]}\left(x,y\right)\right]^{-1}\right\} }{\delta\vartheta_{\mu}\left(z\right)}\label{vertex}
\end{equation}
is the vertex function.

Equations (\ref{D inverso}) and (\ref{S inverso}) show that $\mathcal{D}_{\mu\nu}^{\left[s^{\prime}\right]}\left(x,y\right)$
and $\mathcal{S}_{ab}^{\left[s^{\prime}\right]}\left(x,y\right)$
are the complete Green functions of the theory in thermodynamic equilibrium in the presence of the classical source $J$. When all the sources vanish, we write
\begin{eqnarray}
\mathcal{S}_{ab}^{\left[0\right]}\left(x,y\right) & = & S_{ab}\left(x-y\right)\equiv\mathcal{S}_{ab}\left(x-y\right);\label{Green fermion truly}\\
\mathcal{D}_{\mu\nu}^{\left[0\right]}\left(x,y\right) & = & D_{\mu\nu}\left(x-y\right)-\left\langle \widehat{A}_{\mu}\right\rangle \left\langle \widehat{A}_{\nu}\right\rangle \equiv\mathcal{D}_{\mu\nu}\left(x-y\right).\label{Green photon trully}
\end{eqnarray}

These are the complete Green functions of the physical ensemble. The complete Green functions of the model satisfy the conditions (\ref{S original}) and (\ref{D original}) and are related \textit{via} the Dyson-Schwinger-Fradkin equations (\ref{D inverso},\ref{S inverso},\ref{Pi}-\ref{vertex}) \cite{Dyson DSF,Schwinger DSF,Fradkin DSF}.

We can also investigate how these complete Green Functions in thermodynamic equilibrium behave under a gauge transformation. Let us consider $\widehat{A}_\mu\left(x\right)\rightarrow\widehat{A}'_\mu\left(x\right)=\widehat{A}_\mu\left(x\right)-\partial_\mu\widehat{\varphi}\left(x\right)$, $\widehat{\psi}_a\left(x\right)\rightarrow\widehat{\psi}'_a\left(x\right)=e^{i\widehat{\varphi}\left(x\right)}\widehat{\psi}_a\left(x\right)$, and $\widehat{\overline{\psi}}_a\left(x\right)\rightarrow\widehat{\overline{\psi}}'_a\left(x\right)= \widehat{\overline{\psi}}_a\left(x\right)e^{-i\widehat{\varphi}\left(x\right)}$, where $\widehat{\varphi}$ is an arbitrary self-adjoint $SO\left(2\right)$ scalar field operator. Under this transformation, the Green Functions go to

\begin{eqnarray}
\mathcal{D}_{\mu\nu}\left(x-y\right)\rightarrow\mathcal{D}_{\mu\nu}^{\prime}\left(x-y\right) & = & \mathcal{D}_{\mu\nu}\left(x-y\right)+ \partial_{\mu}^x\partial_{\nu}^y\left\langle T\left[\widehat{\varphi}\left(y\right)\widehat{\varphi}\left(x\right)\right]\right\rangle;\\
\mathcal{S}_{ab}\left(x-y\right)\rightarrow\mathcal{S}_{ab}^{\prime}\left(x-y\right) & = & \mathcal{S}_{ab}\left(x-y\right)\left\langle T\left[e^{-i\widehat{\varphi}\left(y\right)}e^{i\widehat{\varphi}\left(x\right)}\right]\right\rangle.
\end{eqnarray}

These are the so-called Landau-Khalatnikov/Fradkin Transformations \cite{Landau LKF,Fradkin LKF}. This result shows that the Green functions' behavior under a gauge transformation is unaffected by the temperature.

Now, we turn our attention to seek solutions to equations (\ref{psi}-\ref{c barra}). By writing the thermodynamical generating functional as\footnote{Here, $\int_{A-P}D\phi$ means functional integration over anti-periodic fields and $\int_{P}D\phi$ functional integration over periodic fields.}

\begin{eqnarray}
Z_{GF}\left[s\right] & = & \int_{A-P} D\overline{\psi}D\psi \int_{P}DAD\theta D\overline{C}DC\text{ }\widetilde{Z}_{GF}\left[\overline{\psi},\psi,A,\overline{C},C,\theta\right]\times\nonumber \\
&  & \times \exp\left\{\int_{\beta V}d^2z\left[\overline{\eta}_{c}\left(z\right)\psi_{c}\left(z\right) -\overline{\psi}_{c}\left(z\right)\eta_{c}\left(z\right)+J_{\xi}\left(z\right)A_{\xi}\left(z\right) +\varsigma\left(z\right)\theta\left(z\right)\right.\right.\nonumber\\
 & &\left.\left.+\overline{d}\left(z\right)C\left(z\right)-\overline{C}\left(z\right)d\left(z\right)\right]\right\}
\end{eqnarray}
and inserting it into the set of functional equations, we can show that\footnote{We have defined $\kappa\equiv-4i/g$ for simplicity.}

\begin{align}
Z_{GF}\left[s\right] =&\, Z_0 
\left[\det_{P}\left(\Delta+\alpha\frac{q^{2}}{g}\right)\right]^{\frac{1}{2}} e^{\frac{1}{4}\int_{\beta V}d^2rd^2s F\left(z-s\right)\left[\overline{d}\left(r\right)d\left(s\right)-\varsigma\left(r\right)\varsigma\left(s\right)\right]}\nonumber\\
&\,\times Z_{\psi A}\left[J,\eta,\overline{\eta}\right];\label{Z with z}
\end{align}
where $Z_0$ is a (divergent and possibly temperature-dependent) constant, $\det_P$ is the determinant evaluated using periodic functions,
\begin{align}
Z_{\psi A}\left[J,\eta,\overline{\eta}\right] \equiv &\,  \int_{P}DA \int_{A-P}D\overline{\psi}D\psi\, e^{-S\left[\psi,\overline{\psi}, A\right]}e^{\int_{\beta V}d^2z\left[\overline{\eta}_{c}\left(z\right)\psi_{c}\left(z\right)-\overline{\psi}_{c}\left(z\right)\eta_{c}\left(z\right)+ J_{\xi} \left(z\right)A_{\xi} \left(z\right)\right]},\label{chateado}
\end{align}
and $S$ is the thermodynamical effective action:

\begin{align}
S\left[\psi,\overline{\psi}, A\right] \equiv &-\int_{\beta V}d^2z\left\{\overline{\psi}_a\left(z\right)\left(\gamma^E_\mu\right)_{ab}D_\mu^{\left(\mu_f\right)}\left[A\right]\psi_b\left(z\right) +\frac{1}{2}A_\mu\left(z\right) P_{\mu\nu}A_\nu\left(z\right)\right\}.\label{action}
\end{align}

Now, let us perform a gauge transformation in the functional integrals in (\ref{chateado}) accordingly to $A_{\mu}\left(x\right) \rightarrow A_{\mu}^{\prime}\left(x\right)=A_{\mu}\left(x\right)-\partial_{\mu}\zeta\left(x\right)$, $\psi_{a}\left(x\right)  \rightarrow  \psi_{a}^{\prime}\left(x\right)=e^{i\zeta\left(x\right)}\psi_{a}\left(x\right)$, and $\overline{\psi}_{a}\left(x\right)  \rightarrow  \overline{\psi}_{a}^{\prime}\left(x\right)=\overline{\psi}_{a}\left(x\right)e^{-i\zeta\left(x\right)}$, where $\zeta$ is an arbitrary, real $SO\left(2\right)$ scalar field. Under this transformation, $Z_{GF}\left[s\right]$ changes form, yet it remains the same functional. Since the generating functional is originally independent of the gauge function $\zeta$, it must satisfy the condition $\left.\delta Z_{GF}\left[s\right]/\delta \zeta\left(x\right)\right|_{\zeta=0}=0$. This property leads to the Ward-Fradkin-Takahashi identities, written here in the Fourier space \cite{Ward WFT,Fradkin WFT,Takahashi WFT}:



\begin{align}
k_{\mu}^{Bn}\widetilde{\mathcal{D}}_{\mu\nu}^{-1}\left(k^{Bn}\right)=&\, \frac{1}{\alpha q^{2}}\left[\left(k^{Bn}\right)^{2}+\alpha\frac{q^{2}}{g}\right]k_{\nu}^{Bn};\label{Ward identity}\\
p_{\mu}^{Bl}\widetilde{\Gamma}_{\mu\left(ab\right)}\left(k^{Fn},p^{Bl}\right)=&\, \widetilde{\mathcal{S}}_{ab}^{-1}\left(k^{Fn}+p^{Bl}\right)-\widetilde{\mathcal{S}}_{ab}^{-1}\left(k^{Fn}\right),
\end{align}
where $k^{Bn}_1=k^{Fn}_1=k_1$, $k^{Bn}_0=\omega_n^B\equiv 2n\pi/\beta$, and $k^{Fn}_0=\omega_n^F\equiv \left(2n+1\right)\pi/\beta$, for $n$ integer. $\omega^B_n$ ($\omega_n^F$) is called the Bosonic (Fermionic) Matsubara frequency. Actually, due to equations (\ref{P}) and (\ref{D inverso}), the Ward identity (\ref{Ward identity}) implies the transversality to the polarization tensor, \textit{id est} $k^{Bn}_\mu\tilde{\Pi}_{\mu\nu}\left(k^{Bn}\right)=0$, which is a telltale signature of the gauge invariance of the model at quantum level.

\section{The Green Functions and the Partition Function}\label{The Green Functions and the Partition Function}

In this section, we tackle the problem of evaluating the complete two-point Green functions and the partition function of the Gauged Thirring Model.

Upon integration of the antiperiodic Grassmmannian fields in (\ref{chateado}), we find:

\begin{align}
Z_{\psi A}\left[J,\eta,\overline{\eta}\right] = &\, \int_{P}DA\text{ }e^{-S_{A}\left[A\right]}e^{\int_{\beta V}d^2zJ_{\xi}\left(z\right)A_{\xi}\left(z\right)}e^{\int_{\beta V}d^2rd^2s\,\overline{\eta}_{c}\left(r\right)K_{cd}^{\left[\mu_{f}\right]}\left(r,s;A\right)\eta_{d}\left(s\right)}\nonumber\\
&\,\times\mbox{Det}_{A-P}\left\{\gamma^{E}\cdot D^{\left(\mu_{f}\right)}\left[A\right]\right\},\label{later}
\end{align}
where $S_{A}\left[A\right]$ is obtained from the thermodynamical effective action (\ref{action}) by setting the Grassmmannian fields to zero, the capital letter $D$ in $\mbox{Det}_{A-P}$ means the determinant is computed over both the space of functions and the Dirac indices alike and the ``$A-P$" part means the determinant is taken over antiperiodic functions, and $K^{\left[\mu_{f}\right]}\left(\cdot,\cdot;A\right)$ is the Green Function for the operator $\gamma^{E}\cdot D^{\left(\mu_{f}\right)}\left[A\right]$.

Strictly speaking, the determinant of the operator $\gamma_{\mu}^{E}D_{\mu}^{\left(\mu_{f}\right)}\left[A\right]$ is ill-defined when written carelessly and na\"{i}vely. In order to find a meaningful expression for the Green functions and the partition function of the model, we redefine that determinant using the point-splitting regularization \cite{Schwinger}

\begin{eqnarray}
\frac{\mbox{Det}_{A-P}\left\{ \gamma^{E}\cdot D^{\left(\mu_{f}\right)}\left[ A\right]\right\} }{\mbox{Det}_{A-P}\left[\gamma^{E}\cdot\partial^{\left(\mu_{f}\right)}\right]} & = & e^{L\left[A\right]},
\end{eqnarray}
where
\begin{eqnarray}
L\left[A\right] & \equiv & i\int_{\beta V}d^{2}xA_{\mu}\left(x\right)\int_{0}^{1}d\lambda^{\prime}\left(\gamma_{\mu}^{E}\right)_{ab}\lim_{x\leftrightarrow y} K_{ba}^{\left[\mu_{f}\right]}\left(x,y;\lambda^{\prime}A\right) e^{i\lambda^{\prime}\int_{y}^{x}d\xi_{\sigma}A_{\sigma}\left(\xi\right)}e^{\mu_{f}\left(\tau_{x}-\tau_{y}\right)}\label{closed loop functional}
\end{eqnarray}
is the so-called closed-loop functional \cite{Fried}. As it is well known, the limit $x\leftrightarrow y$ must be taken symmetrically.\footnote{In order to avoid any confusion, we will explicitly state what we mean by \textit{symmetrical} limit. The first step is exchanging the variables accordingly to $x_\mu\Rightarrow x_\mu+\epsilon_\mu$ and $y_\mu\Rightarrow x_\mu-\epsilon_\mu$. Then, we write $\epsilon_0=\varepsilon\cos\theta$ and $\epsilon_1=\varepsilon\sin\theta$ and redefine every function $f$ as $f\left(\epsilon\right)\Rightarrow\int_0^{2\pi}f\left(\varepsilon\right)d\theta/2\pi$ before taking the one-sided limit $\varepsilon\rightarrow0^+$.} The presence of the term $e^{\mu_{f}\left(\tau_{x}-\tau_{y}\right)}$ is the subject of a detailed exposition by Alvarez-Estrada and Nicola in \cite{Alvarez Estrada Nicola}. This term only appears when the chemical potential is not zero and, among other features, it ensures the existence of the symmetrical limit. Finding the Green function for the operator $\gamma^{E}\cdot D^{\left(\mu_{f}\right)}\left[ A\right]$ is done through use of the Schwinger \textit{\"{A}nsatz} \cite{Schwinger}

\begin{equation}
K_{ab}^{\left[\mu_f\right]}\left(x,y;A\right)=\left\{ e^{-i\left[\phi\left(x\right)-\phi\left(y\right)\right]}\right\} _{ac}\mathcal{S}_{F\left(cb\right)}^{\left[\mu_f\right]}\left(x-y\right).
\end{equation}
where $\mathcal{S}_{F}^{\left[\mu_f\right]}$ is the Green function for the free operator $\gamma_{\mu}^{E}\partial_{\mu}^{\left(\mu_{f}\right)}$. From here, it follows
\begin{eqnarray}
\left(\gamma_{\mu}^{E}\right)_{ac}\partial_{\mu}\phi_{cb}\left(x\right) & = & \left(\gamma_{\mu}^{E}\right)_{ab}A_{\mu}\left(x\right),
\end{eqnarray}
whose solution is

\begin{eqnarray}
\phi_{ab}\left(x\right) & = & -\int_{\beta V}d^{2}yH_{\left(0\right)}\left(x-y\right)\left(\gamma_{\mu}^{E}\gamma_{\nu}^{E}\right)_{ab}\partial_{\mu}A_{\nu}\left(y\right),
\end{eqnarray}
where $H_{\left(0\right)}$ is the Green Function for the operator $\Delta$ with $\tilde{H}_{\left(0\right)}\left(p^{Bn}\right)=1/\left(p^{Bn}\right)^2$ for its Fourier transform. Therefore, we find for the closed-loop functional:

\begin{eqnarray}
L\left[A\right] & = & -\frac{1}{2\pi}\int_{\beta V}d^{2}xd^{2}yA_{\mu}\left(x\right)\left[ \delta_{\mu\nu}\delta\left(x_{1}-y_{1}\right)\Delta_{\beta}^{\left(+\right)}\left(\tau_{x}-\tau_{y}\right) +\partial_{\mu}^{x}H_{\left(0\right)}\left(x-y\right)\partial_{\nu}^{y}\right] A_{\nu}\left(y\right)+\nonumber \\
 &  & +iF\left(\beta,\mu_f\right) \int_{\beta V}d^{2}xA_{0}\left(x\right),
\end{eqnarray}
with
\begin{align}
F\left(\beta,\mu_f\right)\equiv\frac{\beta\left|\mu_f\right|+\ln\left[\cosh\left(\beta\mu_{f}\right)\right]}{4\pi\beta}.
\end{align}

From these results, we can write

\begin{eqnarray}
Z_{\psi A}\left[J,\eta,\overline{\eta}\right] & = & \mbox{Det}_{A-P}\left[\gamma^{E}\cdot\partial^{\left(\mu_{f}\right)}\right] \int_{P}DA\text{ }e^{\int_{\beta V}d^2rd^2s\overline{\eta}_{c}\left(r\right)K^{\left[\mu_f\right]}_{cd}\left(r,s;A\right)\eta_{c}\left(s\right)} e^{\int_{\beta V}d^2zJ_{\xi}\left(z\right)A_{\xi}\left(z\right)} \nonumber \\
 &  & \times e^{iF\left(\beta,\mu_{f}\right)\int_{\beta V}d^{2}wA_{0}\left(w\right)}e^{-\frac{1}{2}\int_{\beta V}d^{2}xd^{2}yA_{\mu}\left(x\right)W_{\mu\nu}\left(x,y\right)A_{\nu}\left(y\right)}\label{with fermionic sources}
\end{eqnarray}
where
\begin{eqnarray}
W_{\mu\nu}\left(x,y\right) & \equiv & \delta\left(x_{1}-y_{1}\right)\Delta^{\left(+\right)}_\beta\left(\tau_{x}-\tau_{y}\right)\left[\left(\frac{1}{q^{2}}\Delta^{y} +\frac{1}{g}+\frac{1}{\pi}\right)\delta_{\mu\nu}+\frac{1}{q^{2}}\left(1-\frac{1}{\alpha}\right)\partial_{\mu}^{y}\partial_{\nu}^{y}\right]+\nonumber \\
 &  & +\frac{1}{\pi}\partial_{\mu}^{x}H_{\left(0\right)}\left(x-y\right)\partial_{\nu}^{y}.
\end{eqnarray}

The right-hand side of equation (\ref{with fermionic sources}) is hard to be integrated due to the presence of the Green function for the complete Fermionic operator. However, in the case in which the Grassmmannian classical sources vanish, we can rewrite it as

\begin{eqnarray}
Z_{\psi A}\left[J,0,0\right] & = & \mbox{Det}_{A-P}\left[\gamma^{E}\cdot\partial^{\left(\mu_{f}\right)}\right]\left[\mbox{Det}_{P}\left(W\right)\right]^{-\frac{1}{2}}\nonumber\\ &&\,\times e^{\frac{1}{2}\int_{\beta V}d^{2}xd^{2}y\left[J_{\mu}\left(x\right)+iF\left(\beta,\mu_{f}\right)\delta_{0\mu}\right]W_{\mu\nu}^{-1}\left(x,y\right) \left[J_{\nu}\left(y\right)+iF\left(\beta,\mu_{f}\right)\delta_{\nu0}\right]}.\label{almost last}
\end{eqnarray}

Now, the complete electromagnetic Green functions follows immediately from equations (\ref{Green photon}, \ref{Green photon trully}, \ref{Z with z}, \ref{almost last}) and the results from the appendix:

\begin{align}
\mathcal{D}_{\mu\nu}\left(x\right)=&\,\frac{1}{2\pi\beta}\int_{-\infty}^{+\infty}dk_1\sum_{n=-\infty}^{+\infty}\tilde{\mathcal{D}}_{\mu\nu}\left(k^{Bn}\right) e^{ik^{Bn}\cdot x};\\
\tilde{\mathcal{D}}_{\mu\nu}\left(k^{Bn}\right)=&\,\left[\frac{q^{2}}{\left(k^{Bn}\right)^{2}+\frac{q^{2}}{g}+\frac{q^{2}}{\pi}}\right]\left\{ \delta_{\mu\nu}+\left[\frac{\left(k^{Bn}\right)^{2}\left(\alpha-1\right)+\frac{\alpha q^{2}}{\pi}}{\left(k^{Bn}\right)^{2}+\frac{\alpha q^{2}}{g}}\right]\frac{k_{\mu}^{Bn}k_{\nu}^{Bn}}{\left(k^{Bn}\right)^{2}}\right\}.\label{Complete Green Function D}
\end{align}

From this and from (\ref{D inverso}), we find the polarization tensor, whose components in Fourier space are

\begin{align}
\tilde{\Pi}_{\mu\nu}\left(k^{Bn}\right)=\frac{1}{\pi}\left[\delta_{\mu\nu}-\frac{k^{Bn}_\mu k^{Bn}_\nu}{\left(k^{Bn}\right)^2}\right],
\end{align}
from where its transversality can be checked explicitly.

The computation of the complete Fermionic Green function, on the other hand, is more involving. Firstly, we rewrite (\ref{chateado}) as

\begin{eqnarray}
Z_{\psi A}\left[J,\eta,\overline{\eta}\right] & = & \mbox{Det}_{A-P}\left[\gamma^E\cdot \partial^{\left(\mu_f\right)}\right] \left[\mbox{Det}_P\left(P\right)\right]^{-1} e^{i\int_{\beta V}d^2z\left(\gamma_{\mu}^{E}\right)_{ab}\frac{\delta^{3}}{\delta J_{\mu}\left(z\right)\delta\eta_{a}\left(z\right)\delta\bar{\eta}_{b}\left(z\right)}}\nonumber\\
&&\times e^{\int_{\beta V}d^{2}xd^{2}y\overline{\eta}_{c}\left(x\right)S_{F\left(cd\right)}^{\left[\mu_f\right]}\left(x-y\right)\eta_{d}\left(y\right)} e^{\frac{1}{2}\int_{\beta V}d^{2}rd^{2}sJ_{\nu}\left(r\right)P_{\nu\rho}^{-1}\left(r-s\right)J_{\rho}\left(s\right)},
\end{eqnarray}
where $P$ is the differential operator (\ref{P}). Now, using results presented in \cite{Fried}, we can write this as

\begin{eqnarray}
Z_{\psi A}\left[J,\eta,\overline{\eta}\right]
 & = & \mbox{Det}_{A-P}\left[\gamma^{E}\cdot\partial^{\left(\mu_{f}\right)}\right]\left[\mbox{Det}_{P}\left(P\right)\right]^{-1}\nonumber\\
 &  & \times e^{\int_{\beta V}d^{2}xd^{2}y\overline{\eta}_{a}\left(x\right)K_{ab}^{\left[\mu_f\right]}\left(x,y;\frac{\delta}{\delta J}\right)\eta_{b}\left(y\right)}e^{L\left[\frac{\delta}{\delta J}\right]}e^{\frac{1}{2}\int_{\beta V}d^{2}zd^{2}wJ_{\mu}\left(z\right)P_{\mu\nu}^{-1}\left(z-w\right)J_{\nu}\left(w\right)},\label{later Z}
\end{eqnarray}
where $L$ is the closed-loop functional (\ref{closed loop functional}). From (\ref{Green fermion},\ref{Green fermion truly},\ref{Z with z},\ref{later Z}) and still following the steps of \cite{Fried}, it is not too difficult to show that the complete Fermionic Green Function in thermodynamic equilibrium takes the form

\begin{eqnarray}
\mathcal{S}_{ab}\left(x\right)
 & = & \exp\left\{ \left(\frac{g^{2}}{g+\pi}\right)H^{R}_{\left(0\right)}\left(x\right)-gH^R_{\left(\frac{\alpha q^{2}}{g}\right)}\left(x\right)  \right.\nonumber \\
 &  & \left. +\left(\frac{g\pi}{g+\pi}\right)\left[H^R_{\left(\frac{q^{2}}{g}+\frac{q^{2}}{\pi}\right)}\left(x\right) -i\gamma_{1}^{E}\gamma_{0}^{E}F\left(\beta,\mu_{f}\right)x_{1}\right]\right\} _{ac}S_{F\left(cb\right)}^{\left[\mu_f\right]}\left(x\right),\label{Complete Green Function S}
\end{eqnarray}
where $H^R_{\left(m^2\right)}\left(x\right)\equiv H_{\left(m^2\right)}\left(x\right)-H_{\left(m^2\right)}\left(0\right)$ and $H_{\left(m^2\right)}$ is the (periodic) Green Function for the operator $\Delta+m^2$.

For completeness, we shall study several limits of the complete Green functions (\ref{Complete Green Function D}) and ({\ref{Complete Green Function S}}).

First of all, we notice that setting the Fermionic chemical potential to be zero and then taking the limit of vanishing temperature in these Green functions furnishes Euclidean versions of the complete Green Functions for the Gauged Thirring Model at zero temperature presented in \cite{Bufalo Strong Limits}.\footnote{There is a slight difference in the convention in the definition of the gauge parameter in that paper, which translates as $\alpha= q^2\xi$.}

Furthermore, we can show that we can obtain the Schwinger Model's two-point functions from these results. By taking the limit $g\rightarrow\infty$ in (\ref{Complete Green Function D}) and (\ref{Complete  Green Function S}), we find:

\begin{align}
\lim_{g\rightarrow\infty}\tilde{\mathcal{D}}_{\mu\nu}\left(k^{Bn}\right) =&\, \left[\frac{q^2}{\left(k^{Bn}\right)^2+\frac{q^2}{\pi}}\right] \left[\delta_{\mu\nu}-\frac{k_\mu^{Bn}k_\nu^{Bn}}{\left(k^{Bn}\right)^2}\right] +\frac{\alpha q^2}{ \left(k^{Bn}\right)^2}\frac{k_\mu^{Bn}k_\nu^{Bn}}{\left(k^{Bn}\right)^2};\label{D Schwinger}\\
\lim_{g\rightarrow\infty}\mathcal{S}_{ab}\left(x\right)
  = &\, \exp\left\{\pi\left[H^R_{\left(\frac{q^{2}}{\pi}\right)}\left(x\right)-H^{R}_{\left(0\right)}\left(x\right) -i\gamma_{1}^{E}\gamma_{0}^{E}F\left(\beta,\mu_{f}\right)x_{1}\right]  \right.\nonumber \\
   & \left. + \alpha q^2 \left[\bar{B}\left(x\right)-\bar{B}\left(0\right)\right]\right\} _{ac}S_{F\left(cb\right)}^{\left[\mu_f\right]}\left(x\right),\label{S Schwinger}
\end{align}
where $\bar{B}$ is the (periodic) Green function of the forth-order differential operator $\Delta^2$. These limits are, when setting the chemical potential to be zero, exactly the results found in \cite{Ruiz Ruiz Alvarez Estrada Thermo} for the two-point functions of the Schwinger model at finite temperature.

Now, bearing in mind there is no gauge field correlation function in the Thirring Model, we take the limit $q\rightarrow\infty$ in (\ref{Complete  Green Function S}) to find

\begin{align}
\lim_{q\rightarrow\infty}\mathcal{S}_{ab}\left(x\right)
 = \exp\left\{ \left(\frac{g}{g+\pi}\right)\left[ gH^{R}_{\left(0\right)}\left(x\right) -i\pi\gamma_{1}^{E}\gamma_{0}^{E}F\left(\beta,\mu_{f}\right)x_{1}\right]\right\} _{ac}S_{F\left(cb\right)}^{\left[\mu_f\right]}\left(x\right),\label{Limit Thirring S}
\end{align}
which is the result found in \cite{Ruiz Ruiz Alvarez Estrada Thermo} for the complete Fermion propagator in the Thirring Model at finite temperature when their arbitrary function $f$ is chosen to be $f\left(k\right)=1+g^2/\left(\pi+g^2\right)$. This is also the finite-temperature version of the Thirring's Fermionic two-point function found as a limit in \cite{Bufalo Strong Limits} when we put the chemical potential to vanish. Furthermore, the chemical potential dependence of (\ref{Limit  Thirring S}) is the same one presented in \cite{Alvarez Estrada Nicola} for the correspondent complete Green function for the Thirring Model in thermodynamic equilibrium.

All these limits indicate that there is a strong evidence that our results are correct, since in the limit of vanishing temperature and chemical potential they reproduce results found previously at zero temperature, and that the thermodynamic equilibrium does not mar the so-called ``strong limits" of the Gauged Thirring Model \cite{Bufalo Strong Limits}.

As a final subject of our analysis we will compute the partition function of the model. From equations (\ref{part func}), (\ref{Z with z}), and (\ref{almost last}), standard finite temperature calculations yield for the logarithm of the partition function of the Gauged Thirring Model:\footnote{In writing this partition function we used a specific representation of the Euclidean Dirac matrices (\ref{Euclidean Dirac}) $\gamma_{0}^{E}  =  \begin{bmatrix}0 & 1\\
1 & 0
\end{bmatrix}$ and $\gamma_{1}^{E}  =  \begin{bmatrix}0 & -i\\
i & 0
\end{bmatrix}$, and also have absorbed all thermodynamic-irrelevant terms into the constant $Z_0$ which was, then, dropped.}

\begin{eqnarray}
\ln\left[ Z_{GF}\left(\beta,\mu_f\right)\right]  & = & \ln\left(Z_{F}^{0}\right)+\ln\left(Z_{B}^{\frac{q^{2}}{g}+\frac{q^{2}}{\pi}}\right)-\frac{\left[F\left(T,\mu_{f}\right)\right]^{2}}{2}\left(\frac{g\pi\beta V}{g+\pi}\right),\label{partition}
\end{eqnarray}
where
\begin{eqnarray}
\ln\left(Z_{F}^{0}\right) & = & \frac{2V}{\pi}\intop_{0}^{+\infty}dk_{1}\left[\frac{\beta}{2}k+\ln\left(1+e^{-\beta k}\right)\right]
=\frac{2V}{\pi}\intop_{0}^{+\infty}dk_{1}\frac{\beta}{2}k+\frac{\pi V}{6\beta};\\
\ln\left(Z_{B}^{m^{2}}\right) & = & -\frac{V}{\pi}\intop_{0}^{+\infty}dk\left[\frac{\beta}{2}\sqrt{k^{2}+m^{2}}+\ln\left(1-e^{-\beta\sqrt{k^{2}+m^{2}}}\right)\right].\label{Bosonic field}
\end{eqnarray}

Firstly, we notice that, despite the thermodynamical generating functional (\ref{Z with z}) being dependent on the determinant of the differential operator $\Delta+\alpha q^2/g$ - in which the gauge parameter $\alpha$ appears explicitly, the partition function itself is gauge independent and, as a consequence, so are all thermodynamic quantities. In addition, we see that the logarithm of the partition function for the model comprises a sum of three terms: one corresponds to a free, massless Fermionic field $Z_F^0$, one associated with a free, massive Bosonic field $Z_B^{\frac{q^2}{g}+\frac{q^2}{\pi}}$, and a term related to the chemical potential. So, as far as thermodynamics are concerned, all the effects of the interaction between the various fields of the models result uniquely in the coalescence of the mass $m=\sqrt{\frac{q^{2}}{g}+\frac{q^{2}}{\pi}}$ for the Bosonic field. Unfortunately, as far as we know, there is no known, closed form in terms of elementary functions for the last integral in (\ref{Bosonic field}), due to the presence of this very mass. We notice, however, that this mass term possesses the expected behavior when we try to recover the Schwinger and Thirring models from the present model: $m\rightarrow q/\sqrt{\pi}$ as $g\rightarrow\infty$ and $m\rightarrow\infty$ as $q\rightarrow\infty$, respectively \cite{Rosevaldo}. This last limit might seem a little odd at first but, as $\lim_{m\rightarrow\infty}\ln\left(1-e^{-\beta\sqrt{k^{2}+m^{2}}}\right)=0$, we see that it reproduces the correct partition function for the Thirring model computed in \cite{Ruiz Ruiz Alvarez Estrada Thermo}. The term connected to the chemical potential in (\ref{partition}) is interesting because it acts like a completely independent term of the rest of the partition function. Besides, this term remains unaltered in the Thirring Model's limit and goes to $-\pi\beta V\left[F\left(T,\mu_f\right)\right]^2/2$ in the Schwinger Model's. $\ln\left(Z_F^0\right)$, being related to a free field, survives both limits.

\section{Final Remarks}\label{Final Remarks}

In this paper we studied the Gauged Thirring Model in thermodynamic equilibrium. Our approach was based on the Matsubara-Fradkin-Nakanishi Formalism which lies heavily on the density matrix of the grand-canonical ensemble and maintains the Lorentz (or $SO\left(2\right)$, in the Euclidean case) covariance intact during the process of quantization. Using the Schwinger's Principle, we have found the field equations for the model and, from them, we have written the set of functional equations to be satisfied by the thermodynamical generating functional. We studied some general properties of the two-point Green functions of the problem as well, including their symmetries and periodicities, their mutual relationship through the Dyson-Schwinger-Fradkin equations, and their behavior under gauge transformations (the Landau-Khalatnikov/Fradkin transformations). We have seen that all these properties (except, of course, the periodicities) are formally similar to their zero-temperature counterparts. By seeking a solution for the set of functional equations, we have showed the Ward-Fradkin-Takahashi identities satisfied in this model and verified (both through the identities and through explicit computation) the transversality of the polarization tensor.

We have, also, found expressions for the complete two-point Green functions of the Gauged Thirring Model in thermodynamic equilibrium. The electromagnetic Green function acquires a mass $m=\sqrt{\frac{q^{2}}{g}+\frac{q^{2}}{\pi}}$ in the process of quantization in thermodynamic equilibrium and, just like what happens with the Schwinger and Thirring models, this mass is temperature independent. The Fermionic Green function, by its turn, is explicitly dependent on the chemical potential, a result that has shown up previously in the Schwinger Model \cite{Alvarez Estrada Nicola}. It is instructive to recall that if we take the limits of vanishing temperature and chemical potential, both complete Green functions computed in this paper become Euclidean versions of the ones found at zero temperature in \cite{Bufalo Strong Limits}. Since those propagators reproduce the Schwinger's and Thirring's ones at zero temperature when the appropriated ``strong limits" are performed, we can say that, from our results, we can recover those Green functions, too (or, at least without a Wick rotation, Euclidean versions of those propagators). On the other hand, as we have done in this work, we can take the appropriated limits of $g\rightarrow\infty$ and $q\rightarrow\infty$ directly in the complete Green functions in thermodynamic equilibrium. Doing so yields, respectively, in the correspondent Green functions for the Schwinger Model and for the Thirring Model at finite temperature and chemical potential. So, in a sense, our results are rather general: from them we can recover the Green functions for both zero and finite temperature and chemical potential for the Gauged Thirring Model, the Schwinger Model, and the Thirring Model.

Lastly, we computed the partition function for the Gauged Thirring Model. We have shown that, as expected, the partition function is independent of the gauge parameter. This means all thermodynamic quantities are also gauge independent, as it is imperative for physical quantities. The partition function is a product of three terms: one associated with a free, massless Fermionic field, one associated with a free, massive Bosonic field, and a term containing all the chemical potential dependence. The mass of the Bosonic field is the same mass of the gauge field Green function, namely, $m=\sqrt{\frac{q^{2}}{g}+\frac{q^{2}}{\pi}}$. We have shown that both Schwinger Model and Thirring Model's partition functions are ``strong limits" of the Gauged Thirring Model partition function, further corroborating the generality of our results. So, among the three solvable models alluded in this paper, the Gauged Thirring Model in thermodynamic equilibrium is the ultimate one, from which results for the other two can be drawn, being them at zero or finite temperature.

\section{Acknowledgments}

%

CAB thanks FAPESP for partial support and IFT/UNESP, as well, for its hospitality. BMP thanks CNPq for partial support.    

\appendix

\section{A two-dimensional, symmetric, invertible, rank-two tensor in thermodynamic equilibrium}

In this appendix we will consider the most general form for the inverse of a invertible, symmetrical, rank-two tensor in two dimensions in thermodynamic equilbrium with the form

\begin{align}
T_{\mu\nu}\left(k^{Bn}, u\right)=A\delta_{\mu\nu}+B\frac{k^{Bn}_\mu k^{Bn}_\mu}{\left(k^{Bn}\right)^2},
\end{align}
where $u$ is the (Euclidean, two-)velocity of the medium (in the present case, a plasma in 1+1 dimensions), and $A$ and $B$ are two $SO\left(2\right)$ scalar functions that depend on the momentum $k^{Bn}$ and the medium velocity $u$. By \textit{symmetrical} we mean that $T$ satisfies

\begin{eqnarray}
T_{\mu\nu}\left(k^{Bn},u\right) & = & T_{\nu\mu}\left(-k^{Bn},-u\right).\label{sym}
\end{eqnarray}

Its inverse $T^{-1}$ clearly satisfies

\begin{align}
T^{-1}_{\mu\rho}\left(k^{Bn}, u\right)T_{\rho\nu}\left(k^{Bn}, u\right)= T_{\mu\rho}\left(k^{Bn}, u\right)T^{-1}_{\rho\nu}\left(k^{Bn}, u\right)=\delta_{\mu\nu}.\label{inv}
\end{align}

In order to build a rank-two tensor in two dimensions in thermodynamic equilibrium we have at our disposal the objects $k^{Bn}_\mu$, $u_\mu$, and the antisymmetric rank-two tensor $\epsilon_{\mu\nu}$. So, the most general form for $T^{-1}$ is

\begin{eqnarray}
T_{\mu\nu}^{-1}\left(k^{Bn}, u\right) & = & C\delta_{\mu\nu}+D\frac{k^{Bn}_{\mu}k^{Bn}_{\nu}}{\left(k^{Bn}\right)^{2}}+E\frac{k^{Bn}_{\mu}u_{\nu}}{k^{Bn}\cdot u}+F\frac{u_{\mu}k^{Bn}_{\nu}}{k^{Bn}\cdot u}+G\frac{\left(k^{Bn}\right)^{2}u_{\mu}u_{\nu}}{\left(k^{Bn}\cdot u\right)^{2}}+H\varepsilon_{\mu\nu}\nonumber\\
&&+I\frac{\bar{k}^{Bn}_{\mu}k^{Bn}_{\nu}}{\left(k^{Bn}\right)^{2}}+J\frac{k^{Bn}_{\mu}\bar{k}^{Bn}_{\nu}}{\left(k^{Bn}\right)^{2}} +K\frac{\bar{k}^{Bn}_{\mu}\bar{k}^{Bn}_{\nu}}{\left(k^{Bn}\right)^{2}}
 +L\frac{\bar{k}^{Bn}_{\mu}u_{\nu}}{k^{Bn}\cdot u}+M\frac{k^{Bn}_{\mu}\bar{u}_{\nu}}{k^{Bn}\cdot u}\nonumber\\
 &&+N\frac{\bar{k}^{Bn}_{\mu}\bar{u}_{\nu}}{k^{Bn}\cdot u}+O\frac{\bar{u}_{\mu}k^{Bn}_{\nu}}{k^{Bn}\cdot u}+P\frac{u_{\mu}\bar{k}^{Bn}_{\nu}}{k^{Bn}\cdot u}+Q\frac{\bar{u}_{\mu}\bar{k}^{Bn}_{\nu}}{k^{Bn}\cdot u}+R\frac{\left(k^{Bn}\right)^{2}\bar{u}_{\mu}u_{\nu}}{\left(k^{Bn}\cdot u\right)^{2}}\nonumber\\
 &&+S\frac{\left(k^{Bn}\right)^{2}u_{\mu}\bar{u}_{\nu}}{\left(k^{Bn}\cdot u\right)^{2}}+U\frac{\left(k^{Bn}\right)^{2}\bar{u}_{\mu}\bar{u}_{\nu}}{\left(k^{Bn}\cdot u\right)^{2}},
\end{eqnarray}
where $C$, $D$, ..., $U$ are coefficients with similar properties to $A$ and $B$ and, for any $SO\left(2\right)$ vector $a$, we wrote $\bar{a}_{\mu}  \equiv  a_{\nu}\varepsilon_{\nu\mu}$.

Due to equations (\ref{sym}) and (\ref{inv}), we find

\begin{eqnarray}
E=F=G=H=I=J=K=L=M=N=O=P=Q=R=S=U=0
\end{eqnarray}
and
\begin{eqnarray}
1 & = & AC\\
0 & = & AD+BC+BD
\end{eqnarray}

So, provided the system above has a solution, the most general form for the inverse of the tensor $T$ is
\begin{eqnarray}
T_{\mu\nu}^{-1}\left(k^{Bn}, u\right) & = & C\left(k^{Bn}, u\right)\delta_{\mu\nu}+D\left(k^{Bn}, u\right)\frac{k^{Bn}_{\mu}k^{Bn}_{\nu}}{\left(k^{Bn}\right)^{2}}.
\end{eqnarray}


\end{document}